**This is the accepted manuscript (postprint) of the following article:**
N. Zirak, A. Maadani, E. Salahinejad, N. Abbasnezhad, M. Shirinbayan, *Fabrication, drug delivery kinetics and cell viability assay of PLGA-coated vancomycin-loaded silicate porous microspheres*, Ceramics International, 48 (2022) 48-54.
https://doi.org/10.1016/j.ceramint.2021.08.298

# Fabrication, drug delivery kinetics and cell viability assay of PLGA-coated vancomycin-loaded silicate porous microspheres

N. Zirak [a], A.M. Maadani [a], E. Salahinejad[*,a], N. Abbasnezhad [b,c], M. Shirinbayan [b,c]

[a] Faculty of Materials Science and Engineering, K. N. Toosi University of Technology, Tehran, Iran

[b] Arts et Metiers Institute of Technology, CNAM, PIMM, HESAM University, Paris, France

[c] Arts et Metiers Institute of Technology, CNAM, LIFSE, HESAM University, F-75013, Paris, France

## Abstract

Porous ceramic microspheres are a desirable substance for bone tissue reconstruction and delivery applications. This study focuses on Mg-Ca silicate microspheres encapsulated in biodegradable poly (lactic-co-glycolic acid) (PLGA) to serve as a biocompatible carrier for the controlled release of vancomycin hydrochloride. In this regard, diopside ($MgCaSi_2O_6$), bredigite ($MgCa_7Si_4O_{16}$), and akermanite ($MgCa_2Si_2O_7$) powders were synthesized by sol-gel and subsequent calcination methods. Then, porous akermanite, diopside and bredigite microspheres of 700-1000 μm in diameter were fabricated by using carbon porogen, droplet extrusion and sintering, then loaded with the drug and eventually coated with PLGA. The bare microspheres showed a considerable burst release mode of the drug into a physiological medium, whereas

[*] Corresponding Author: Email Address: <salahinejad@kntu.ac.ir>

PLGA coating of the microspheres reduced the burst release level. To investigate effective mechanisms governing in the drug release from the carriers, the contribution of burst, degradation, and diffusion was analyzed by the sequential quadratic programming algorithm method. It was found that the relative contribution of diffusion to bioresorption is ranked as diopside > akermanite > bredigite, whereas PLGA coating dominates the diffusion mechanism. The dental pulp stem cells cytocompatibility MTT assay of the microspheres also showed that the drug loading deteriorates but PLGA coating improves the cell biocompatibility significantly. Comparatively, the biocompatibility of the PLGA-coated microspheres was ranked as akermanite > diopside > bredigite, as a result of a compromise between the release of the constituting ions of the ceramic carriers and vancomycin molecules. It was eventually concluded that PLGA-coated Mg-Ca silicate microspheres are promising candidates for drug-delivery bone tissue engineering and dental bone grafting applications.

## 1. Introduction

Trauma, congenital deformity and pathological deformation are the main causes of bone defects, which can be effectively treated by using bone substitutes and grafts [1, 2]. Among 3D scaffolds used for this purpose, microspheres benefit from several advantages, including suitable shape and flowability for filling complex defects as well as drug loading ability [3]. Porosity inside these carriers also plays a key role in providing proper cell adhesion, the transport of nutrients, growth factors and oxygen and in determining the drug loading and delivery [4]. In addition, empty spaces created between the microspheres after implantation can act as appropriate sites for bone and vascular ingrowth [5, 6]. As an ideal bone substitute, some bioceramics from the $SiO_2$-CaO-MgO system have attracted considerable attention owing to suitable biocompatibility, bioresorbability and bioactivity [7, 8]. The most typical members of this system are bredigite ($Ca_7MgSi_4O_{16}$), diopside ($CaMgSi_2O_6$) and akermanite ($Ca_2MgSi_2O_7$) which have been successfully used in different shapes including microsphere [9-11].

Due to the significance of antibacterial properties and dealing with infection after the implantation of microspheres, vancomycin was normally loaded in these substances as an antibiotic drug. Achieving a suitable drug release profile is critical for healing and has been one of the most important goals in many studies. It has been reported that bare Mg-Ca silicate microspheres provide a burst release of drugs which disadvantageously affects medication therapy and biocompatibility [9-11], as the problem of this research. One of the most common methods to obtain the suitable properties is biopolymer coating of the carriers. Polycaprolactone (PCL) and poly (lactic-co-glycolic-acid) (PLGA) are biodegradable polymeric candidates for encapsulation. Excellent biocompatibility, successful controlled release of drugs, minimal inflammatory response during degradation are the main advantages of these two polymers [12-14]. Considering the less degradation rate of PCL compared to PLGA [15], PCL encapsulation may suppress the useful bioresorbability of akermanite, diopside and bredigite. Thus, the use of PLGA encapsulation was hypothesized in this research to control the drug delivery of the microspheres.

To our knowledge, the effect of polymeric coating on the bioperformance of vancomycin-loaded Mg-Ca silicate microspheres has not been reported to date. Diopside, bredigite and akermanite porous microspheres were accordingly produced by a droplet extrusion route, impregnated with vancomycin hydrochloride and encapsulated in PLGA in this work. Then, the drug release from the carriers was analyzed under static conditions to determine the delivery kinetics. Finally, the cytocompatibility of the microspheres was investigated by the MTT assay.

## 2. Experimental procedures

### *2.1. Fabrication and structural characterization of microspheres*

Bredigite ($MgCa_7Si_4O_{16}$), diopside ($MgCaSi_2O_6$) and akermanite ($MgCa_2Si_2O_7$) powders were synthesized by sol-gel techniques using calcium nitrate tetrahydrate ($Ca(NO_3)_2{\cdot}4H_2O$, Merck, >98%), magnesium nitrate hexahydrate ($Mg(NO_3)_2{\cdot}6H_2O$, Merck, >98%) and tetraethyl orthosilicate ($(C_2H_5O)_4Si$, TEOS, Merck, >98%) precursors. In summary, TEOS was first hydrolyzed in 2 M nitric acid aqueous solution for 30 min. Then, the stoichiometric values of $Ca(NO_3)_2{\cdot}4H_2O$ and $Mg(NO_3)_2{\cdot}6H_2O$ were added and stirred. The achieved solution was aged and gelated at 60 °C for 24 h and then dried at 120 °C for 2 days. The obtained xerogels were calcinated at 1300 °C.

Microspheres were produced from the calcined silicate powders, based on Ref. [10]. In brief, the 2:3 mixture of the synthesized ceramic and spherical carbon (Sigma Aldrich, >99.95%) was added to a sodium alginate ($NaC_6H_7O_6$, Sigma Aldrich) solution (3 wt%). Afterward, the slurry was dropwise added into a 0.1 M calcium chloride ($CaCl_2$, Merck, >98%) solution, leading to the precipitation of microspheres which were then followed by washing with distilled water several times. The obtained microspheres were dried at 60 °C, sintered at 1250 °C, and observed by scanning electron microscopy (SEM, HITACHI 4800).

### *2.2. Loading of microspheres with drug and encapsulation in PLGA*

The microspheres were loaded with vancomycin hydrochloride ($C_{66}H_{75}Cl_2N_9O_{24}$, Sigma Aldrich) according to Ref. [16]. Briefly, one gram of each microsphere was placed in a 10 ml tube with a 1 mm diameter hole at the end. The tube was placed upside down at the end of a 50 ml round-bottomed flask. After the evacuation of the flask and tube, a solution of vancomycin and water at the concentration of 10 mg/ml from stopcock entered the flask under evacuation. The vacuum valve was then opened and the drug solution was introduced into the tube under air pressure. The microspheres and drug solutions were kept for 24 h, then passed through a filter and dried at 60 °C.

To control the drug release from the microspheres, the impregnated carriers were coated with PLGA (5004 A, Corbion, Netherlands). For this purpose, one gram of the microparticles was immersed in 5% w/v PLGA-acetone solution for 30 sec. Finally, after filtering the microspheres, they were air-dried for one day.

### *2.3. Drug delivery studies of microspheres*

To investigate the drug release behavior, 1 g of the impregnated microparticles was soaked in 10 ml of phosphate buffered saline (PBS) at 37 °C. Sampling was performed at 1, 3, 6, 9, 24, 48, 72, 96, 120, 144 and 168 h, so that 4 ml of new PBS was exchanged at each time. The drug concentration was determined by a UV spectrophotometer (UV–1100, MAPADA INSTRUMENT, Shanghai, China) with at least three repetitions. Furthermore, the carriers were ultrasonicated for 30 min after 168 h of immersion in PBS to estimate the total amount of the drug loaded in the samples.

### *2.4. Cytocompatibility assessment of microspheres*

To study the biocompatibility of the samples, three milligrams of the microparticles were placed in a 1.5 ml microtube and sterilized in 100 µl ethanol followed by 2 h of UV radiation. 3500 dental pulp stem cells were then seeded on a 96-well culture plate. The microparticles were added to the culture cell environment with three replicates for each test. The medium was then added to the wells and stored at 37 °C for different days at an incubator. The MTT protocol was then performed on the samples to determine the cell viability using an ELISA reader. The suitability of dental pulp stem cells for bone-related studies has been previously pointed out in the literature [17-19].

## 3. Results and discussion

### *3.1. Structure characterization*

Figure 1 depicts the macrograph and micrograph of the uncoated microspheres. Based on the images, the diameter of the sintered microspheres is in the range of 700-1000 μm. The micrographs also confirm the presence of micropores of 10-120 μm in size inside the microspheres due to the carbon porogen burning during sintering, acting as the matrix and path of the drug incorporation and release from the inside out. Ideal pore diameters inside and between bone void fillers for cell seeding, tissue growth, vascularization, nutrient delivery and waste removal are in the range of 10–500 μm [20, 21]. Furthermore, the critical size of micropores on the fillers for osteointegration is in the range of 2–10 μm [22, 23], reflecting the desired geometrical characteristics of the produced microspheres. Regarding the phase analysis of the produced microspheres, X-ray diffraction has previously verified the merit of the same processing methods used in this study for synthesizing the desired crystalline phases of bredigite, diopside, and akermanite [9].

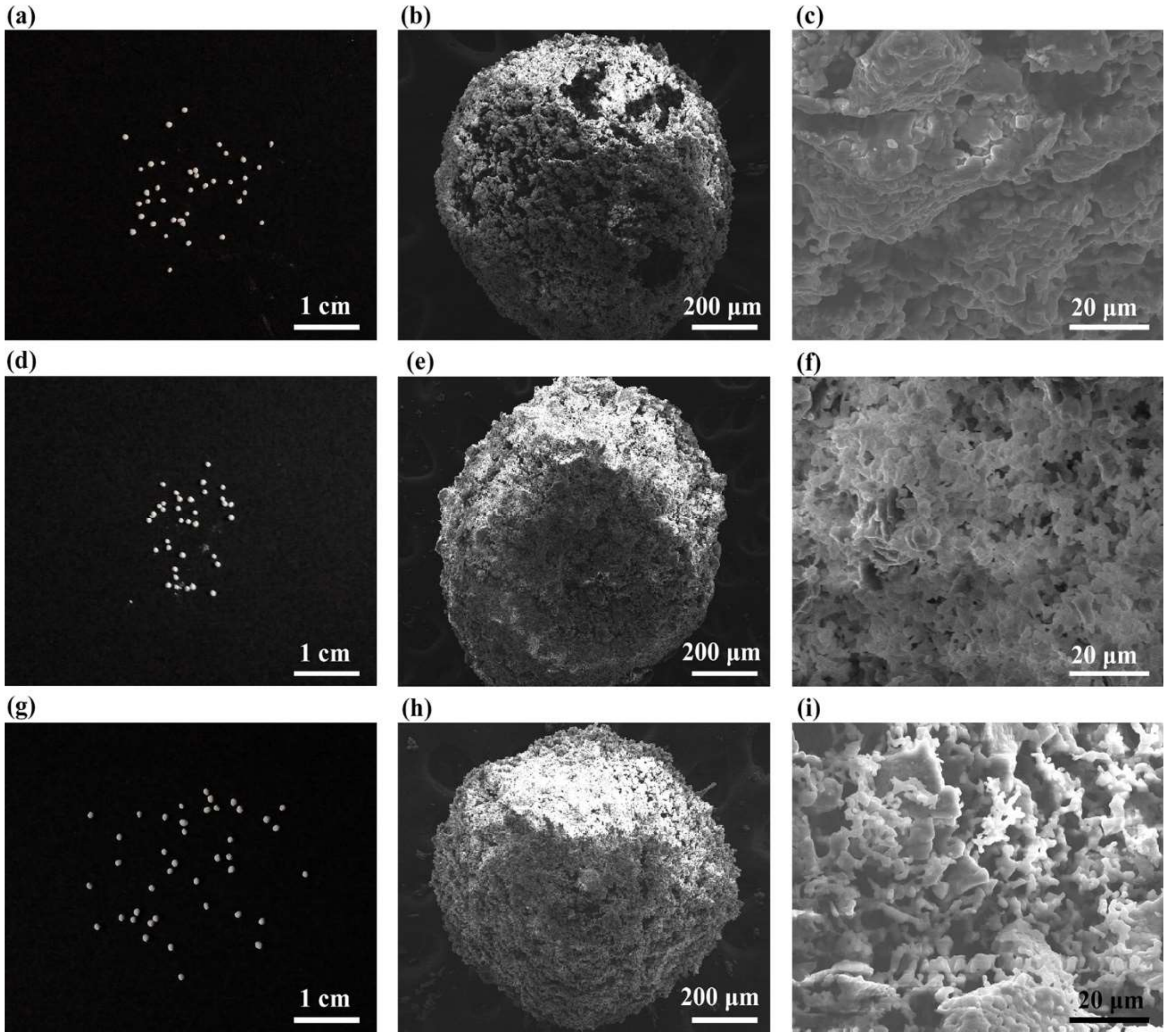

**Figure 1.** Macrograph and SEM micrograph of the bare bredigite (a, b, c), akermanite (d, e, f), and diopside (g, h, i) microspheres.

Figure 2 indicates the SEM micrograph of the akermanite, bredigite and diopside microspheres encapsulated in PLGA. By coating the microspheres with the polymer, a decrease is observed in the surface roughness of the microspheres compared to the bare microspheres. However, the PLGA concentration of the coating polymeric solution has not been so high that the deposited PLGA layer is able to fill pores, producing a rough coating on the surfaces. The presence of these surface micropores on the coated microspheres is beneficial for cell adhesion and proliferation required for bone tissue regeneration [24].

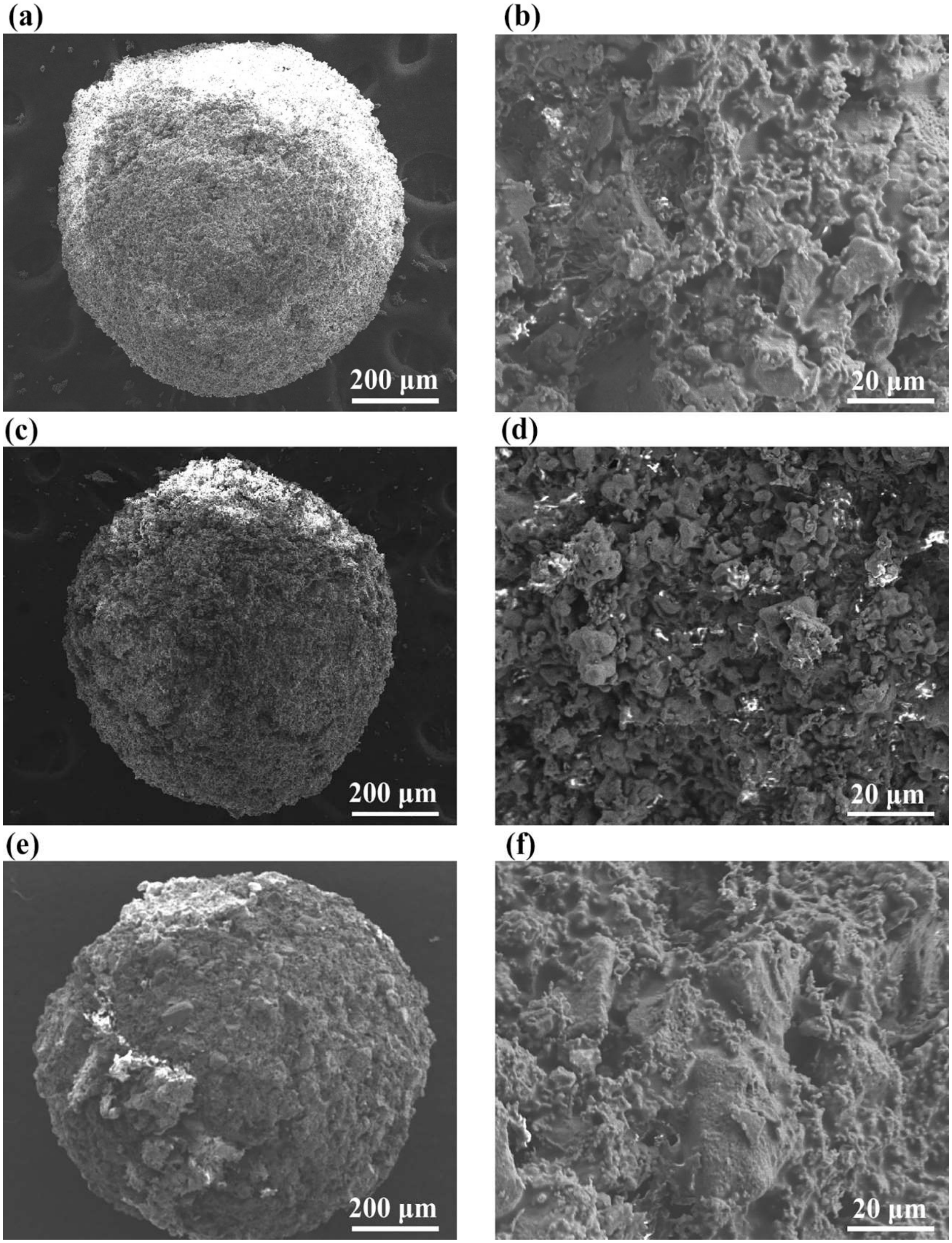
(a)
200 μm
(b)
20 μm
(c)
200 μm
(d)
20 μm
(e)
200 μm
(f)
20 μm

**Figure 2.** SEM micrograph of the PLGA-coated bredigite (a, b), akermanite (c, d), and diopside (e, f) microspheres.

### *3.2. Experimental drug release studies*

The concentration and cumulative level of the drug release from the microspheres are shown in Figure 3. The total amount of the drug loaded into one gram of the bredigite, diopside, and akermanite microspheres was estimated to be 15.9 ± 0.2, 16.6 ± 0.1 and 17.3 ± 0.1 mg, respectively. According to Figure 3(a) and (b), the bare microspheres indicated a burst drug release mode in the first 9 h of exposure, and then a sustained release. The use of the PLGA coating limits the burst release from all the microspheres, as shown in Figures 3(c) and (d). For each group of the bare and coated samples, the drug release rate was ranked as bredigite > akermanite > and diopside carriers. Typically, the bredigite microspheres show 54% of the total drug incorporated at the initial burst release, which is reduced to 41% after using the polymer coating. The level of burst release from the akermanite microspheres is 41% for the uncoated state and 36% for the coated state. The lowest release rate was obtained for the diopside microspheres, which was 31% of release for the uncoated condition and 30% for the coated state. That is, by decreasing the drug release rate from bredigite to akermanite and to diopside, the effect of the PLGA coating on the control of burst release is also reduced. Indeed, the enhancement in the concentration of Si and thereby bridging oxygen (Si-O-Si) in these Mg-Ca silicates from bredigite ($MgCa_7Si_4O_{16}$) to akermanite ($MgCa_2Si_2O_7$) and then to diopside ($MgCaSi_2O_6$) is responsible for differences in bioresorption rate and hence drug release rate [25].

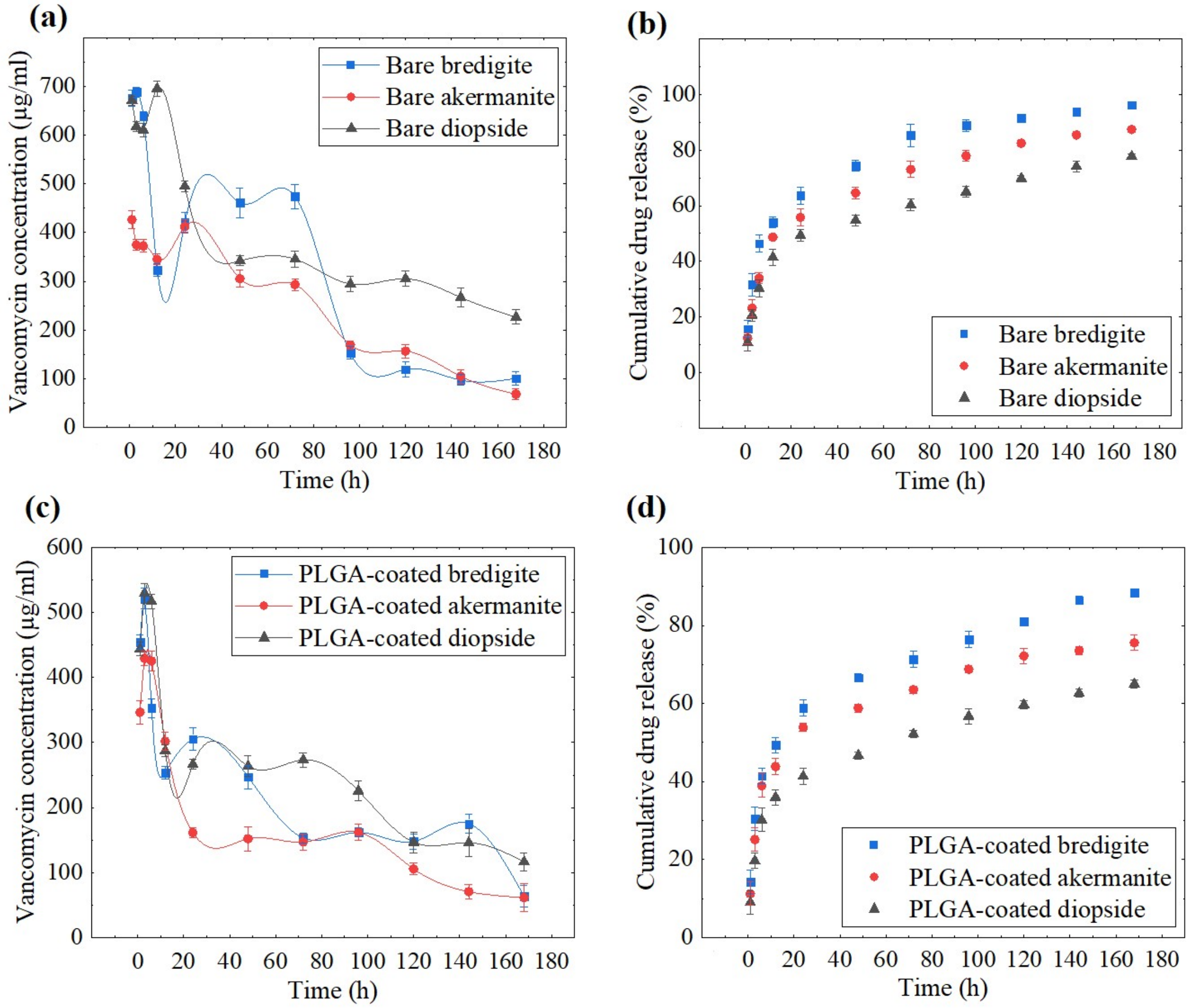

**Figure 3.** Concentration and cumulative amount of vancomycin released from the bare (a, b) and PLGA-coated (c, d) microspheres, respectively.

To evaluate the merit of the drug delivery vehicles prepared in this work with respect to the therapeutic window, the level of the drug released during the whole study time (Figures 3(a) and (c)) is considered. The treatment of osteomyelitis demands vancomycin release continuously at concentrations beyond 0.75 µg/ml [26, 27]. As can be seen, for all the samples and over the total drug release period, the vancomycin level is beyond the critical value of treatment, which reflects the ability of these microspheres to inhibit the bacterial growth.

### *3.3. Drug delivery kinetics evaluations*

Kinetics-based mathematical models play an important role in improving the design of drug-delivery systems. In general, these models concern with drug release mechanisms, according to the properties of the carriers [28]. In this regard, the critical role of some mechanisms like burst release has received a lot of attention, especially for antibiotic therapy because the burst release of some antibiotics in the first 6 h of implantation is beneficial to prevent the adhesion of bacteria [29, 30]. It has been shown that the release of drugs from silicate ceramics is according to diffusion and bioresorption or a combination of these two mechanisms. In addition, the effect of polymer coating on drug release mechanisms from bredigite scaffolds loaded with vancomycin encapsulated in PLGA has been previously investigated [31]. In this section, considering mathematical models, the contribution of the drug release mechanisms from the microsphere carriers is analyzed.

Non-linear equation systems that model the burst, diffusion-controlled and degradation-controlled release of drugs were solved by a sequential quadratic programming algorithm (SQPA) in Matlab software in this study, as it is recognized to be the most effective optimization procedure for solving non-linear optimization problems [32, 33]. For the examination of these phenomena, the following equations with especial coefficients were used:

$$\frac{M_t}{M_\infty} = \sum_{i=1}^{i=N} \mu_i \times F_i \quad (1)$$

where $M_t$ shows the drug level release at time *t*, $M_\infty$ is the drug level release at the infinitive time, the contribution of each of the N mechanisms is examined by $\mu_i$ where $0 \leq \mu_i \leq 1$ and $\sum_{i=1}^{i=N} \mu_i = 1$, and $F_i$ is the corresponding equations.

The burst release of drugs is defined as [34]:

$$\frac{dC}{dt} = -k_b C \quad (2)$$

where $k_b$ is the burst release constant and *C* represents the drug concentration in the media at time *t*. Considering the fact that the drug concentration in the microsphere equals to the preliminary weight of the solute per the volume of the microsphere, the following equation is accordingly obtained [34, 35]. Accordingly, the following model provides $F_1$ for the burst release in Equation 1:

$$F_1 = \frac{M_t}{M_\infty} = 1 - \exp(-k_b t) \quad (3)$$

The Baker and Lonsdale's model [36] was used in this study to describe the diffusion mechanism ($F_2$). This model is a modification of the Higuchi model for spherical, heterogeneous particles, as follows:

$$F_2 = \frac{3}{2}\left[1 - \left(1 - \frac{M_t}{M_\infty}\right)^{2/3}\right] - \frac{M_t}{M_\infty} = kt \quad (4)$$

where *k* is a drug release constant.

The Hixson–Crowell model [37] which is based on degradation was also used for analyzing the drug release from the degradable/resorbable microspheres ($F_3$). The model is written as below:

$$F_3 = \sqrt[3]{1 - \frac{M_t}{M_\infty}} = -kt \tag{5}$$

The final equation for the drug release in these systems was obtained as below:

$$\frac{M_t}{M_\infty} = (\mu_1 \times F_1) + (\mu_2 \times F_2) + (\mu_3 \times F_3) \tag{6}$$

where the values of $\mu_1$, $\mu_2$, and $\mu_3$ represent the contribution of the burst, diffusion, and degradation phenomena, respectively.

The combined model developed for the prepared drug delivery systems (Equation 6) was then fitted to the experimental results of vancomycin released from the akermanite, diopside and bredigite microspheres (Figure 4) and the fitting parameters were listed in Tables 1 and 2. The considerable values of *R*, correlation coefficient, for all of the systems represent the excellent fitting of the data. For the bare state, the estimated contribution of the burst release for the diopside and akermanite microspheres shows an insignificant difference, whereas this mechanism for the bredigite sample is meaningfully dominated. Also, the resorption-to-diffusion mechanisms ratio us ranked as: bredigite > akermanite > diopside. This is compatible with the study of Wu et al. [25] concluding the same ranking for the bioresorption rate of these Mg-Ca biosilicates. Comparing Tables 1 and 2, it can be observed that the contribution of the burst mechanism for all the carriers is reduced as a result of PLGA coating. Regarding the competitive contribution of diffusion and resorption/degradation among the different coated carriers, the ranking is similar to the bare conditions, i.e. the increase in the degradation-to-diffusion mechanisms ratio from diopside to akermanite and to bredigite. However, the degradation-to-diffusion mechanisms ratio for all the microspheres is decreased by using PLGA encapsulation, due to the fact that vancomycin release from PLGA is mainly diffusion-controlled [31].

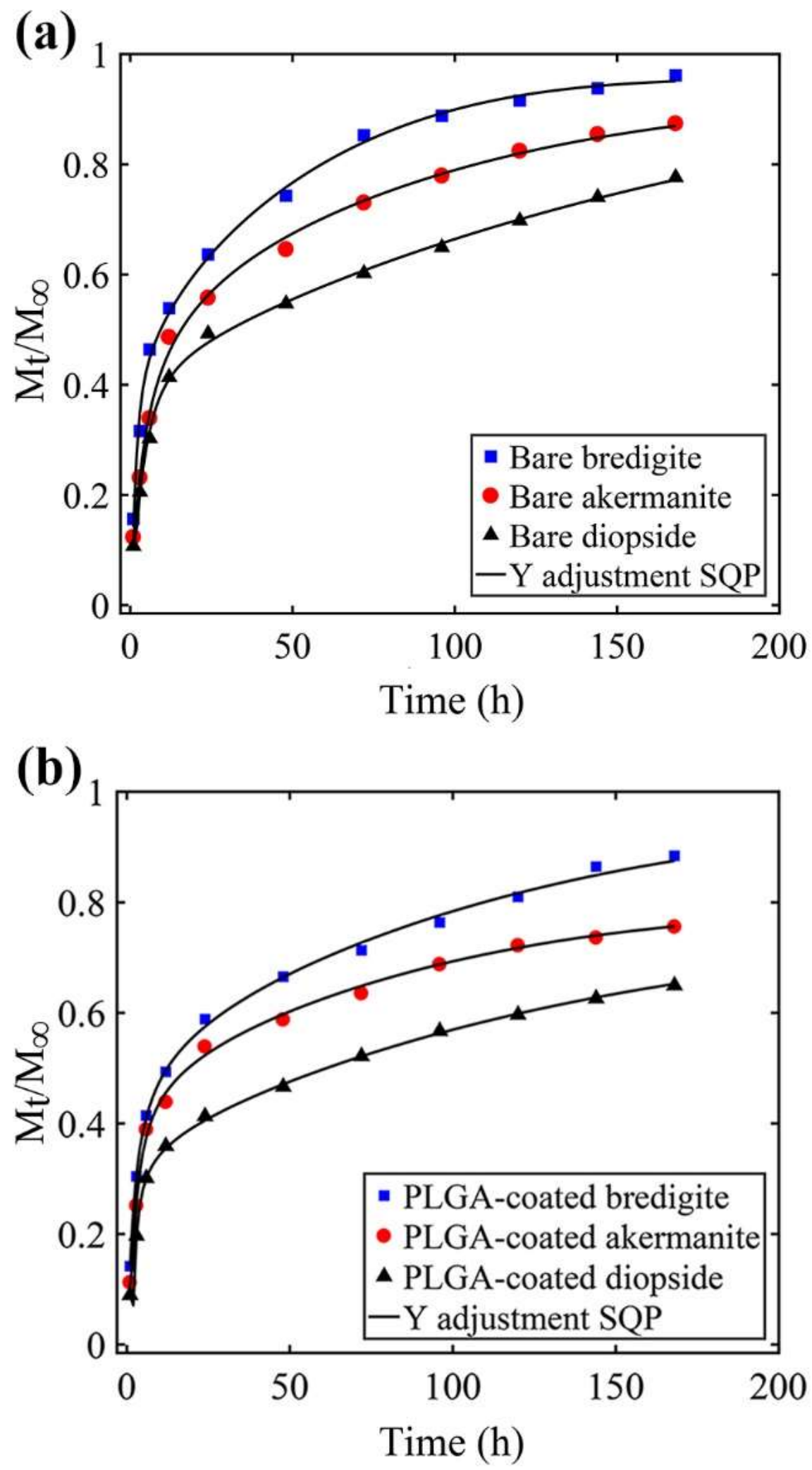

**Figure 4.** SQPA model fitting curves of vancomycin hydrochloride released from the bare (a) and PLGA-coated (b) microspheres.

**Table 1.** Contribution level of vancomycin hydrochloride release mechanisms from the bare microspheres.

| Sample | Burst (%) | Diffusion (%) | Resorption (%) | Resorption/Diffusion | R |
|---|---|---|---|---|---|
| Bredigite | 49 | 17 | 34 | 2.0 | 0.99 |
| Akermanite | 47 | 37 | 16 | 0.4 | 0.98 |
| Diopside | 38 | 50 | 12 | 0.24 | 0.99 |

**Table 2.** Contribution level of vancomycin hydrochloride release mechanisms from the coated microspheres.

| Sample | Burst (%) | Diffusion (%) | Degradation (%) | Degradation/Diffusion | R |
|---|---|---|---|---|---|
| Bredigite | 42 | 21 | 37 | 1.8 | 0.99 |
| Akermanite | 43 | 45 | 12 | 0.3 | 0.98 |
| Diopside | 31 | 60 | 9 | 0.15 | 0.98 |

### *3.4. Cell biocompatibility analyses*

The MTT testing results of dental pulp stem cells culture on the samples are depicted in Figure 5. A considerable cell proliferation for all the samples is realized from the increase in the optical density (OD) of viable cells with culture time from the $1^{st}$ to the $7^{th}$ day. The cell viability on all of the three drug-free microspheres at all the culture periods shows no statistically significant difference with the control, confirming the cell biocompatibility of the synthesized Mg-Ca silicate microspheres. However, the impregnation of the microspheres with vancomycin has meaningfully reduced the cell cytocompatibility. Coating of the vancomycin-loaded microspheres with PLGA has caused a better cell viability, even slightly better than the vancomycin-free samples.

The cell proliferation is encouraged by the bioresorption of akermanite, diopside and bredigite and subsequently the beneficial release of Mg, Ca and Si ions from the bioceramics toward the culture environment. It is well-established that due to the contributions of the silicon level and bridging oxygen, the bioresorption rate of these Mg-Ca silicates is ranked as bredigite > akermanite > diopside [25]. Additionally, the continuous release of the ceramic constituents, i.e. Mg, Ca and Si ions when exposed in physiological environments has been previously pointed out diopside [38, 39], akermanite [40, 41] and bredigite [42, 43]. Nonetheless, comparing the cell viability on the three types of the carriers, it is inferred that bredigite in all of the drug-free, drug-loaded and PLGA-coated states exhibits the lowest cytocompatibility than akermanite and diopside. This is due to the higher bioresorbability of this ceramic, causing the higher release of calcium ions and hence metabolic alkalosis effect disadvantageously [44]. In this regard, akermanite demonstrates the highest cell viability since it benefits from an optimal rate of bioresorption. The improvement in biocompatibility due to PLGA coating of the microspheres can be related to the alteration of the burst drug release in the bare conditions into the more sustained release of vancomycin. Also, the higher viability on the coated carriers in comparison to the drug-free microspheres suggests that the PLGA coating controls not only vancomycin release but also the excess bioresorption rate of the bioceramics. Eventually, the akermanite microspheres are concluded to be the optimal choice in all the drug-free, drug-loaded and polymer-encapsulated conditions for bone void filling applications.

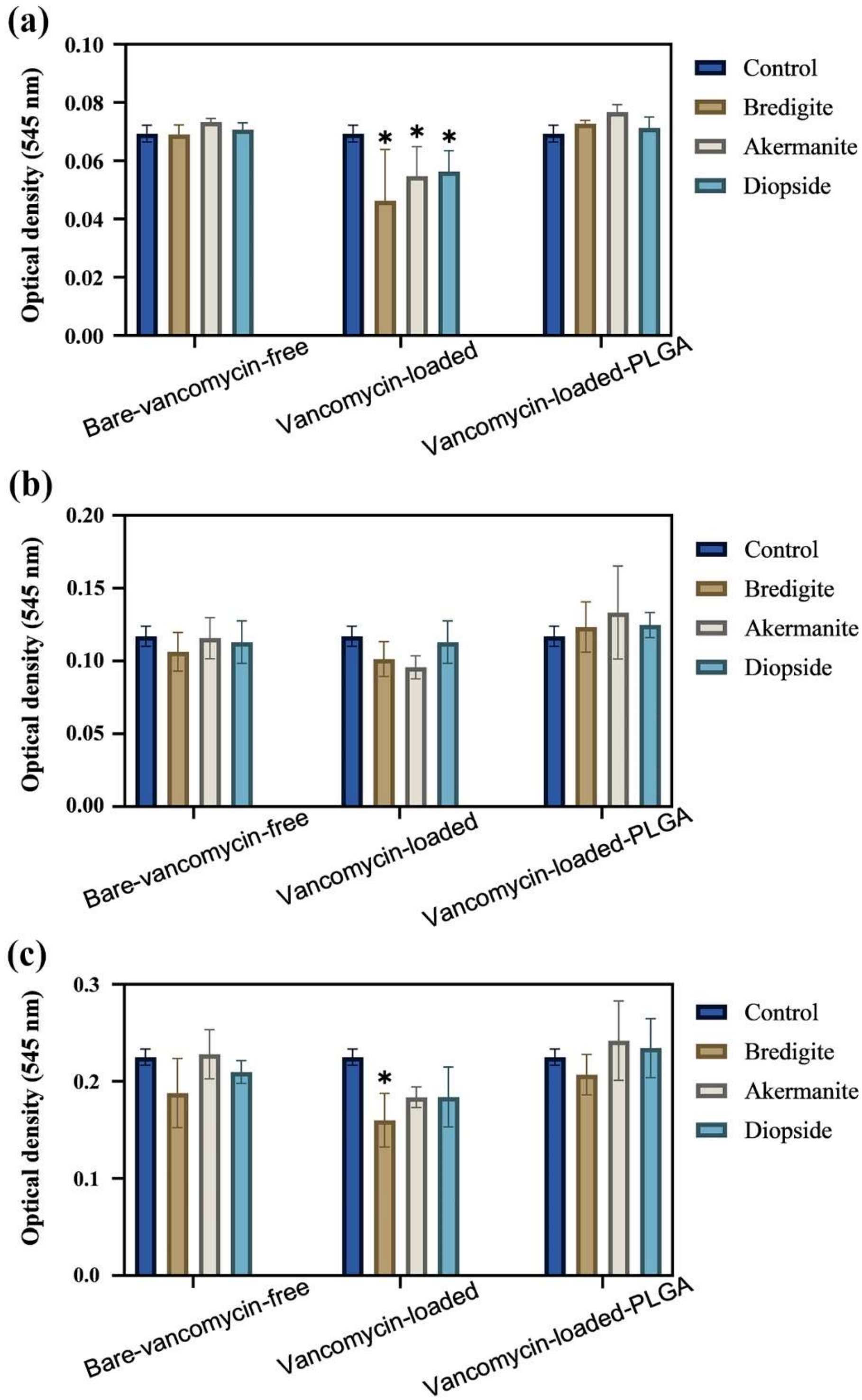
(a)
Optical density (545 nm)
0.10
0.08
0.06
0.04
0.02
0.00
*
*
*
Bare-vancomycin-free
Vancomycin-loaded
Vancomycin-loaded-PLGA
Control
Bredigite
Akermanite
Diopside
(b)
Optical density (545 nm)
0.20
0.15
0.10
0.05
0.00
Bare-vancomycin-free
Vancomycin-loaded
Vancomycin-loaded-PLGA
Control
Bredigite
Akermanite
Diopside
(c)
Optical density (545 nm)
0.3
0.2
0.1
0.0
*
Bare-vancomycin-free
Vancomycin-loaded
Vancomycin-loaded-PLGA
Control
Bredigite
Akermanite
Diopside

**Figure 5.** Results of dental pulp stem cell culture for the different periods: 1 (a), 3 (b), and 7 (c) days. ∗ demonstrates significant differences in comparison to the control with $P < 0.05$.

## 4. Conclusions

In this study, porous akermanite, diopside and bredigite microspheres were prepared by sol-gel, calcination, droplet extruding and sintering methods. The microspheres were loaded with vancomycin hydrochloride and then coated with PLGA to optimize the drug release kinetics and biocompatibility. The typical achieved results are as follows:

1. Microspheres with desirable dimensional characteristics, including the diameter of 700-1000 µm and the open pore size of 10-120 µm, were successfully synthesized by the used processes.
2. The uncoated microspheres showed a considerable burst release step of vancomycin due to the domination of the bioresorption mechanism, so that the contribution of this mechanism was ranked as bredigite > akermanite > diopside with a reverse order for the diffusion release mechanism.
3. The encapsulation of the microspheres with PLGA limited the burst release of the drug as a result of the increase in the role of the diffusion mechanism compared to the bare states.
4. The drug release kinetics indicated that during the whole test time, the concentration of vancomycin was above the critical value for stopping staphylococcus aureus growth. This suggests the merit of the prepared drug delivery microspheres with respect to the therapeutic window.

5. The akermanite microspheres coated with PLGA showed the highest cell viability with respect to dental pulp stem cells. The findings suggested that the effect of the drug release kinetics on cell cytocompatibility is higher than that of the bioresorption rate of the biosilicates carriers.